# THE WORK OF ART IN AN
# AGE OF MECHANICAL GENERATION

Steven J. Frank

In 2017 a cultural milestone was reached: an artificial intelligence (AI) system generated a series of artworks that the public couldn't distinguish, in terms of origin, from works created by recognized (human) artists.[1] The randomized, double-blind study was billed as a "visual Turing test" after the famous gauge of conversational machine intelligence. Subjects were fooled at least 75% of the time.

But in fact, this was no Turing test. It asked random people with expertise neither in art nor in AI to make a gut-level judgment based on unstated criteria. Many preferred the AI-generated art to that created by recognized artists. This tells us that the computer-generated artwork pleased the eye and projected enough aesthetic sophistication to function as art. But what does it mean to "function" as art? There was no real engagement between the subjects of the study and the artwork they viewed because their task was to answer a question, not to interact meaningfully with a creative work. And that's true of any casual interaction with art, whether on the way to a law firm's conference room or driving past a mural. Art "functions" in such glancing encounters if it confirms the prestige and tasteful judgment of your lawyer and no one drives off the road.

Art museum and gallery visitors, in contrast, have different expectations because they arrive with purpose. They willingly submit themselves to the curatorial authority that has selected *this* art for display on high-rent walls, and in return for our devotions we expect something from the artist that transcends mere decorative merit. We study the work and consider the artist and her intent. What influenced her? Can a social context be perceived or an insight gained? Is there an explicit rejection of earlier work or its subversive re-use?

Perhaps a Turing test can be constructed from this deeper experience of viewing art and contemplating its human dimensions. We might ask our test subjects whether, as they view an image, they can perceive in the work some of these indicia of intentionality — whether the subject matter imparts enough creative force to convey a unity of medium and message, maybe even a story or prophetic warning. Of course, much visual art is concerned solely with form and great comedic art may turn on the pretentious twaddle of a bourgeois gentilhomme. But let's focus on art that, at some level, combines form with meaning and give our volunteer the benefit of the doubt. Have we found a test for the human in a work of art?

Maybe not. Because, first, recent AI-generated artworks are pretty good, and it's actually not hard to imagine, at some level, ideas that could have influenced and be evident in their form. And second, because every human element mentioned thus far can be — in fact has been — coded into AI systems. At the foundational level of aesthetics, the study of what pleases the human eye, psychologists and their forbears have identified key stimuli such as novelty, complexity, surprise, and incongruity. Similar criteria inform a neuroscientific definition of creativity.

AI systems called "generative adversarial networks" (GANs) synthesize images, eval-

---

uate them against some criterion, adjust the image to reduce the deviation, and repeat the process until some performance measure is satisfied. If the criterion is an elephant, the GAN takes digital hammer and chisel and carves away everything that isn't an elephant. But the criterion can be anything, including one or more aesthetic stimuli. A GAN can digest the entire canon of Western art and evaluate synthesized images for dissimilarity or similarity to any or all of the canonical works, producing a novel image as easily as an homage. Human-created artworks can be scored for the presence of attributes associated with originality and lasting influence, and the GAN, in its automated pas de deux between creator and critic, will embed those attributes in its output. It can scan the web for social context and add that, too. And over time, its performance will improve in ways that defy prediction.

So awed were the ancient Greeks by the notion of creativity that they attributed it to not one but several goddesses. Yet the more we understand what it means to be creative (even if we may disagree over the fundamental attributes), the more actionable criteria we can supply to an AI. Even intentionality can be faked if we can identify enough examples. You search in vain for the quintessentially human but it turns out there's an app for that.

Or is there?

### SIGNATURES AS SIGNIFIERS

Which is more valuable — a lock of Abraham Lincoln's hair or his autograph? Each is a direct, tangible link to a revered historical figure. Yet a cursory web search suggests they sell for roughly comparable prices (in five figures) despite the relative scarcity of the former.

The lock of hair is a holy relic of a secular saint. The tradition of venerating the body parts and clothing of worthy figures is an ancient one that, even today, stirs considerable emotion and attracts crowds of the faithful and the curious. When Notre Dame burned in April 2019, news stories often led with the headline that the cathedral's Crown of Thorns, reputed to have been placed on Christ's head by Roman soldiers, had survived the blaze. Catholic tradition prescribes a hierarchical classification: first-class relics include the bones or other body parts of a saint; the clothing or other objects used by the saint, such as rosaries, are second-class relics, while articles that have come into contact with a first- or second-class relic thereby qualify as third-class relics. Earlier systems of belief explicitly ascribed magical properties to the remains of kings and other demigods.

Religious significance aside, the lock of hair is dead, a remnant — not materially different from any other lock of hair. Even Lincoln's DNA is just DNA. Survive though it may as a chemical residue of Lincoln's flesh, it carries nothing of Lincoln as "prairie lawyer," leader during his country's most convulsive conflict, and national martyr. A signature, to be sure, is also just chemicals, dark stains on a pulp-based substrate. But its connection to Lincoln transcends chemistry.

A signature is a trace grounded in volition, a stamp of human consciousness. The signer of any document knows that his mark signifies more than authentication or assent; it will be perceived as a personal statement, and in an earlier day of pen and letters, great efforts were made to display originality and style. John Hancock's flamboyant signature on the Declaration of Independence evokes the defiant stroke of a bold man before tyranny, and even if the truth is more prosaic — he seems to have always signed his name that way — it's hard not to read it as conveying something of the man's manner and style. The pseudoscience of graphology goes a step further in viewing handwriting features as manifestations of specific personality characteristics. Long ago debunked, graphology none-theless retains its adherents, one of whom recently offered this psychoanalysis of Lincoln through his handwriting:

> "He has what is called print writing. His writing is a combination of someone who prints and cursives all in one. You find that in very, very intelligent people's handwriting. They combine the most efficient things of cursive and printing. These are your very, very creative thinkers. … There's no pretense. 'I'm a simple man. I'm not here to put up this big front for everybody else.'"[2]

Many who would scoff at the magical powers of a bone fragment might nonetheless credit the signature as a mystical window to the soul. And even those who reject graphology altogether must recognize the signature as a purposeful, communicative gesture captured permanently in ink. As such, a genuine signature is valuable because it serves as a medium of communion that no chemical residue can approach. It not only encodes humanity but imparts it. A signature has voice.

The purchaser of an autograph, then, has acquired an intimate psychic bridge to the signer, and even if this bridge is more imagined than real, it cannot survive discovery that the signature is a fake. Perfect fidelity to the genuine mark is irrelevant; it is the intimacy, not the ink, that underlies value.

### SEEKING THE ELUSIVE SIGNATURE

A painting is a vast collection of brushstroke signatures spread over (at least) two dimensions and extending in time, i.e., the duration of the artist's efforts at the easel. But with rare exceptions, artists apply pencil or paint to a substrate to achieve a final unified image, not to make a personal statement with each mark. The brushstrokes of neoclassical painters such as Jean-Auguste-Dominique Ingres are barely discernible.

If the artist's statement is her finished work rather than her stroke, then the sum has so wholly subsumed the parts as to present a conundrum. A signature's function — to signify a formal act of adoption, assent or authentication — is routine and utilitarian. An artwork may also serve utilitarian ends, as noted earlier, but these have a much broader social context. More important, they are subsidiary to the work's far more imperative aesthetic, decorative, and communicative objectives. From what source, then, does the value of authenticity arise? Your engagement with a Rembrandt portrait may lead you to contemplate the sitter and her times, social distinctions encoded in dress (those gaudy collars!), myriad themes great and small — but not, or certainly not exclusively, the personality of a particular seventeenth-century Dutch painter. If Rembrandt's work is not prized for any psychic link to Rembrandt the man, why does the value of a de-attributed Rembrandt plunge five orders of magnitude when not one brushstroke, and certainly none of the higher-order attributes of aesthetics and message, has changed?

Walter Benjamin incisively explored this question in his famous 1935 essay, "The Work of Art in the Age of Mechanical Reproduction." To Benjamin, a work extends beyond its physical lineaments, which can be reproduced, to include its history and context — its "aura" — which inheres profoundly but invisibly. "A strange weave of space and time," Benjamin wrote of this notion, "the unique appearance or semblance of distance, no matter how close it may be." If invisible, however, the

---

aura's significance must be quasi-religious and grounded in ritual.

It's questionable whether, in an age of triumphant consumerism and declining religiosity, a basis in ritual (and the prestige of ownership) can support the prodigious value discrepancy between the demonstrably and the plausibly genuine. With some predictable exceptions, prices for Old Master paintings and drawings have been falling for years, mirroring broader trends in luxury goods. Whereas luxury once implied bespoke craftsmanship and prices affordable only by a few, today's luxuries — such as Apple's iPhone — are relatively low-priced products with mass appeal. Perhaps we are witnessing a decline in the value of aura as a commodity.

Complicating questions of value for works of art are controversies surrounding attribution. And no artist's work has undergone so tumultuous a saga of expert rejection and rehabilitation as have Rembrandt's. A century ago, his total output was estimated at 711 works. That number began to dwindle, soon quite dramatically, following the establishment of the Rembrandt Research Project (RRP) in 1968. Members of this committee, Dutch art historians charged with the task of de-attributing dubious Rembrandts, frequently disagreed over stylistic criteria, and disagreement itself often resulted in de-attribution. Dozens of works were rejected. By 1989, only 250 works had survived the RRP's judgment. Although the committee later restored 90 or so works to Rembrandt's oeuvre before disbanding in 2011, many paintings remain controversial.

In fairness, Rembrandt works are notoriously difficult to authenticate. Like many European artists of his era, he ran a workshop of close imitators, many with substantial talent. He is believed to have signed, or allowed his signature to appear on, paintings from his workshop that he actually did not paint. His work has frequently been forged.

Although the RRP employed scientific methods such as dendrochronology (which determines the age of a wood panel), textile research, analysis of paint samples, and radiographic techniques, these approaches could not distinguish between works by Rembrandt and the contemporaneous efforts of his students. The RRP eventually fell back on traditional connoisseurship, but concerns arose over "overly strict use of stylistic criteria of authenticity" and "certain a priori assumptions about the (possibly too narrow) limits of variability within Rembrandt's style,"[3] in the words of Rembrandt expert and RRP member Ernst van de Wetering. In short, the RRP found itself torn between objective criteria inapplicable to difficult cases and subjective judgments that settled nothing definitively.

The erudite, mordant, and joyously unrepentant forger Eric Hebborn derided the RPP as a coterie of self-styled "Rembrandt revisionists" whose opinions could not be trusted. Hebborn the forger took particular delight in fooling his share of connoisseurs, lambasting the art expert as "a congenial manipulator both of history and taste, who must have everything just as he wants it. Our job [as forgers] is to satisfy his feelings in the matter." His particular objection to the RRP was its failure to credit what, to Hebborn, was the most definitive evidence of all: "Painters who do not smooth out their brushwork are signing their work with every stroke." Ignore this obvious truth, Hebborn tells us, and "we find ourselves in the absurd position of having to postulate a genius other than Rembrandt who has painted this body of masterworks and join those crackpots of the kind who attribute Shakespeare's works to Marlowe, Bacon and others."[4]

---

*Pace* Hebborn, my own adventures as a computer scientist with an interest in neural networks casts at least some doubt on the notion of brushstroke as fingerprint. A branch of AI, neural networks loosely mimic the brain's organization and, like the brain, excel at recognizing patterns. "Convolutional" neural networks (CNNs) analyze and classify images based on visual content. A CNN starts out dumb but becomes smart as it's trained. In theory you could train a CNN to classify paintings as Rembrandt or not Rembrandt by showing it many, many images labeled as one or the other. But there are a few problems. First, CNNs can only handle small images; an image of an average-sized portrait with sufficient detail to support analysis will be many times too large. Second, a CNN must be trained with thousands of images, and there are only a few hundred confirmed Rembrandts. Finally, even if a CNN could somehow process a large, high-resolution image, it would probably see too much at once to produce a meaningful classification. It's one thing to detect whether a picture contains a dog, quite another to distinguish between a Rembrandt from a Flinck.

My approach has the computer divide a large image into many small, equally sized square tiles and sift the ones with enough pictorial information to, in effect, pinch-hit for the whole. Each tile has as much visual diversity — though of course not visual content — as the entire image. The tiles are small enough for a CNN to process and plentiful enough to train it effectively. To evaluate a candidate image, the system decomposes it into tiles and finds the best ones, and the CNN assigns a Rembrandt/non-Rembrandt probability score to each tile. The average of the scores represents the system's judgment.

We can train the CNN at different tile sizes. Smaller tiles concentrate the analysis on diminutive visual features like brushstrokes; bigger tiles allow consideration of compositional elements. If the "training set" of images is well-chosen, with the non-Rembrandts having deliberately varying degrees of pictorial similarity to the Rembrandts, the CNN will learn to make fine distinctions. With training and test sets of Rembrandt portraits assembled by my wife Andrea Frank, an art historian, we found something interesting and unexpected. Classification accuracy spikes at a certain tile size corresponding, roughly, to a face, and drops off sharply at smaller sizes. At the brushstroke level the CNN is just guessing, with accuracies no better than a coin flip.

A brushstroke is certainly a signature in the sense of a physical trace imparted by the will and volition of a human hand. But our results suggest that it may not uniquely identify its maker. Perhaps this should not be surprising. The application of paint combines motor skill, material properties, and a choice of technique. These factors, ultimately, constrain the possibilities. Rembrandt's students strove to imitate his technique, and the most talented ones succeeded. Harder to duplicate are the more imaginative, dramatic, communicative — the more *human* — elements of Rembrandt's style and work. These are his true signature, and they cannot be characterized or catalogued. Our CNN can discriminate based on them but can't explain what they are. Even if we coax the CNN to show us what regions of a painting contribute most strongly to its classification, we find these to be, unexcitingly, the brightest and busiest areas. The CNN can report where it's looking but not what it sees. The human remains as elusive in the aloof confidence of an AI as it does in the noisy squabbles of connoisseurs over Rembrandt versus Rembrandtesque.

## ART, ARTIST, AND ARTIFICIAL INTELLIGENCE

Prognostications that AI will surpass and eventually replace human-created artwork echo, very obviously, similar warnings issued at the dawn of photography. But for the most part, art absorbed rather than succumbed to the new medium. Recognizing the aesthetic potential of action frozen in time and experience directly recorded, artists, including painters such as Thomas Eakins, embraced the camera as a tool. Eventually the cultural tent stretched to accommodate the formal qualities of a photograph — particularly the sharply focused verisimilitudes of Modernism — as legitimate artistic subject matter. Roughly a century after its invention, photography officially entered the cultural mainstream as a department of the Museum of Modern Art.

But the advent of the camera forever altered the market for and the utilitarian function of painting and drawing. It wiped out the centuries-old trade in portrait miniatures and challenged all forms of representational artwork. Introspection over the unique qualities inherent in painting contributed to the rise of Impressionism and Expressionism (as well as the eventual man-bites-dog reciprocation of Photorealism). Will AI follow a similar trajectory?

Artists have embraced AI as a tool. Christie's held its first auction of AI art in 2018, and one work, the GAN-created *Portrait of Edward Bellamy*, sold for $432,500 — nearly 45 times its already-high estimate. Moreover, if the 20th-century evolution of painting toward abstraction represents, in part, the determination of painters to extend the non-photographic aspects of their medium, AI can be expected to provoke similar efforts to privilege human perception and craft over the mechanical. AI may eventually recruit additive manufacturing technologies such as 3D printing to create a convincing painting, for example, but the robotic equipment required to compose a fabric collage or papier-mâché assemblage will not likely be developed to serve the art market. (Of course, like Sol Lewitt or Jeff Koons, an AI could generate assembly instructions or blueprints.)

AI has the potential to radically revise the relationship among art, artist, and the market. If a GAN can not only hone a work to conform to arbitrarily supplied criteria but also assimilate and respond to language, the need for an artist mediating between medium and consumer disappears — the GAN can make art to order and to specification. While such products may strike some as gimmicky schlock, their proud owners might perceive testaments to their own artistic refinement. Surely some such merchandise could pass the art Turing test. With the convergence of luxury and mass appeal, the possibility of a market for artist-free art isn't unthinkable.

Still, it's probably a mistake to dismiss the importance of art as an expression of human life, its longings and struggles, even in works emphasizing the formal. Ultimately what we want from art, as opposed to our furniture, can never be satisfied solely by checking the aesthetic boxes. But preserving the human dimensions may require more effort from artists to engage their increasingly distracted patrons. As the personal experience of viewing art in an intimate gallery setting yields to online commerce and art fairs that have become celebrity pageants, artists who offer perfunctory statements of intent (which often themselves seem computer-generated) and assume their work will otherwise speak for itself may find their voices increasingly lost in the cacophony. Personal engagement, tweets, blogs, and TED talks enhance both the artist and her work. If art can no longer reliably function as a window to the human, the humans must assertively claim the foreground.

Artists eternally strive to outrace obsolescence and seek new paths to originality. If we're willing to accept their use of technology as an expressive tool, it will enlarge their work rather than replace it. *You cannot devour me*, says the artist to the advancing robot, *for I am omnivorous*.

### The Work of Art in an Age of Mechanical Generation
<u>Bibliography and Further Reading</u>

Space limitations preclude listing supporting references within the text of this article, and doing justice to the rich literature surrounding art, artificial intelligence, aesthetics, and notions of authenticity requires more than spot citations in any case. This supplemental bibliography attempts at least to skim the sea of thought devoted to the subjects covered.

### The Technology

Although invented quite recently,[5] generative adversarial networks (GANs) have received widespread attention, stimulating diverse creative efforts and fears of rogue disinformation campaigns (more on the latter below). The GAN-created *Portrait of Edmond Belamy* that sold for nearly a half-million dollars may have attracted the most attention, but "original" works produced by GANs and their users have become routine, now occupying their own niche in the art market.[6] GANs can use art-historical and social context to create new works[7] and have been used to generate imitations within established art genres, most notably (and recently) Chinese landscape painting.[8] Convincing though the Chinese landscapes may be, they are 512×512 pixel thumbnails, not full-size works.

Convolutional neural networks (CNNs) have a much older pedigree; François Chollet[9] and Aurélien Géron[10] provide good introductions to their technical details. My own work is described in greater detail in two recent articles[11] co-authored with my wife. Prior efforts using CNNs to analyze artwork include Lecoutre et al.,[12] Elgammal et al.,[13] Lyu et al.,[14] and van Noord et al.[15]

All technology has a dark side, and artificial intelligence (AI) is no exception. GANs have given rise to a "deepfakes" industry that, when not amusing us with images of happier Mona Lisas[16] and "pizza that does not exist,"[17] terrifies us with the prospect of made-up news videos and libelous pastiches of the real invisibly stitched into the fake.[18] More broadly, neural networks — like any machine learning tool — reflect the biases of their training datasets and, ultimately, their creators. Whether due to mundane lapses in data hygiene or more insidious motivations, the resulting systems, if used unquestioningly, can cause harm. One study examined commercial facial-analysis software and found an error rate of 0.8% for light-skinned men but 34.7% percent for dark-skinned women.[19] According to the paper, a major U.S. technology company that claimed an accuracy rate of more than 97% assessed performance on a sample that was more than 77% male and more than 83% white. Combating such bias may require greater public participation in scientific discourse[20] and more sophisticated methods of decision making under conditions of uncertainty.[21]

### Aesthetics

The literature examining the psychological origins of aesthetic judgments is substantial; a leading general reference is the *Oxford Handbook of Aesthetics*.[22] The neuroscience of creativity remains an embryonic field; researchers are only beginning to link functional neuroanatomy to higher brain functions, much less to specific, complex psychological states. Keith Sawyer[23] provides a worthwhile critical overview of current research and Arne Dietrich[24] describes an interesting, though conjectural, framework associating creative processes with neural domains. A less technical summary of current understandings is provided by Tim Newman.[25]

The interplay among the biology of visual perception and processing, emotional response, and mathematical relationships inherent in aesthetically pleasing works and phenomena has fascinated thinkers for millennia. Aristotle famously defined beauty as having order, symmetry and definiteness, but it was this journal's namesake, Leonardo da Vinci, who broadly united an aesthetic view of science with the scientific study of aesthetics. In his magisterial biography of Leonardo, Walter Isaacson relates this anecdote:

> While at Windsor Castle looking at the swirling power of the "Deluge drawings" that [Leonardo] made near the end of his life, I asked the curator, Martin Clayton, whether he thought Leonardo had done them as works of art or of science. Even as I spoke, I realized

it was a dumb question. "I do not think that Leonardo would have made that distinction," he replied.[26]

One wonders what Leonardo the humanist would have made of GANs.

Contemporary treatments of aesthetic principles often consider them in light of sociopolitical changes or behavior-influencing advances in technology. Nicolas Bourriaud, for example, considers the effects of interactivity and mass communication on evolving tastes in art, proposing an understanding of art as the exchange of information between artist and audience.[27]

Whether machine creativity is possible or an oxymoron may be an undecidable question, but the challenge is joined in numerous recent works with little consensus emerging. Whereas Arthur I. Miller argues that computers can already be as creative as humans and will someday surpass us,[28] the philosopher Sean Dorrance Kelly sees creativity as exclusively human.[29]

### Authenticity and the Art of Authentication

Walter Benjamin's notion of an artwork's aura reflects his struggle to identify the root value of material art when the cost of material duplication was plummeting. Champions of conceptual art such as Joseph Kosuth, on the other hand, deny the existence of such value altogether, arguing that art only begins where materiality ends.[30]

The eternal arms race between art forgers and those charged with their unmasking is well described by Noah Charney.[31] Charney surveys both scientific and connoisseurship approaches to authentication. Catherine Scallen delves more deeply into the practice of connoisseurship with particular attention to Rembrandt.[32]

In advancing his view of brushstrokes as signature elements of artistic style, Eric Hebborn cites with approval the Scottish chemist A.P. Laurie, author of *The Brushwork of Rembrandt and His School* (1932), who similarly believed that "[t]he drawing with the brush of each painter differs more markedly than men's signatures differ from one another." Although our work challenges this notion for Rembrandt, the same is not true for van Gogh, as we discuss in our 2020 *Neural Computation* article.[33] To a CNN, Van Gogh's uniqueness *only* emerges at the brushstroke level; for purposes of differentiating his work from that of other artists, his subject-matter treatment and compositional arrangements apparently are not unique enough.

---

[26] Walter Isaacson, *Leonardo da Vinci* (Simon & Schuster 2017).
[27] Nicolas Bourriaud, *Relational Aesthetics* (Les Presse Du Reel 1998).
[28] Arthur I. Miller, *The Artist in the Machine* (MIT Press 2019).
[29] Sean Dorrance Kelly, "What computer can't create," *Technology Review*, (2019).
[30] Joseph Kosuth, *Art After Philosophy and After: Collected Writings, 1966–1990* (MIT Press 1991).
[31] Noah Charney, *The Art of Forgery* (2015).
[32] Catherine Scallen, *Rembrandt, Reputation, and the Practice of Connoisseurship* (2004).
[33] Frank and Frank [7].